\documentclass[twocolumn,showpacs,preprintnumbers,amsmath,amssymb]{revtex4}
\usepackage{graphicx}
\usepackage{dcolumn}
\usepackage{bm}
\begin{document}
\title {Failure of the local density approximation in time-dependent \\
spin density functional theory}
\author {Z. Qian, A. Constantinescu, and G. Vignale}
\affiliation
{Department of Physics and Astronomy,  University of Missouri, Columbia,
Missouri
65211}
\date{\today}
\begin{abstract}
It has been known for some time that the exchange-correlation potential
in time-dependent density functional theory is an intrinsically
nonlocal functional of the density as soon as one goes beyond the adiabatic
approximation.  In this paper we show that a much more
severe nonlocality problem, with a completely different physical origin,
plagues the exchange-correlation potentials in time-dependent {\it
spin}-density
functional theory.  We show how the use of the {\it spin current density}
as a basic variable solves this problem,
and we provide an explicit local expression for the exchange-correlation
fields
as functionals of the spin currents.
\end{abstract} \pacs{71.15.Mb, 71.45.Gm, 71.10.Ca}
\maketitle
For many years the local density approximation (LDA) has provided the
much needed handle on the difficult problem of approximating
the density dependence of the exchange-correlation (xc) potential -- the
single particle potential that incorporates the many-body effects in the
Kohn-Sham equation for the ground state density \cite{DreizlerGross}.
In LDA, the xc potential $V_{xc}(\vec r)$  is simply a function
of the local density $n(\vec r)$.  This approximation is not unreasonable
as long as the functional derivative of $V_{xc}(\vec r)$ with
respect to $n(\vec r')$ -- the so called {\it exchange-correlation kernel}
$f_{xc}(\vec r, \vec r') \equiv
\frac {\delta V_{xc}(\vec r)}{\delta n(\vec r')}$ -- is
 a sufficiently short-ranged function of the distance  $|\vec r - \vec r'|$   \cite{footnote1}.  

However, much recent work \cite{Vignale1,Gonze1,Martin,Vanderbilt,Tokatly} has demonstrated
that the  requirement of short-rangedness is not always
fulfilled in physical systems, and when this happens the local
density approximation is flawed.  This does not mean that a local
description of  exchange and correlation is absolutely impossible,
only that such a description cannot be achieved in terms of
the particle density.

For example, in the density-functional theory of crystalline insulators,
it has been found \cite{Gonze1,Martin,Vanderbilt} that
the xc potential has an
``ultranonlocal" dependence on the density, due to the fact that the Fourier transform of the xc
kernel $f_{xc}(\vec k, \vec k)$ diverges as $1/k^2$ for
$k \to 0$ in these systems.
But, the ultranonlocality disappears if one reformulates the theory in
terms of the electric polarization $\vec P(\vec r)$ and the
exchange-correlation electric field $\vec E_{xc} (\vec r )$
associated with it.

A similar phenomenon was discovered in the time-dependent density
functional theory (TDDFT)\cite{Grossdobsonpetersilka}
following the realization that the frequency-dependent LDA\cite{GK} fails to
satisfy Kohn's theorem\cite{Dobson,Kohnstheorem}.  The pathology was traced
to a singularity of the form $\frac{\vec k \cdot \vec k'}{k^2}$ in the xc kernel $f_{xc}(\vec k, \vec k',\omega)$ for $k \to 0$
at  finite $\vec k'$ and $\omega$.  The  ensuing nonlocality problem was
solved by upgrading  to time-dependent {\it current}-density functional
theory (TDCDFT),  where the basic variable is the current density, and its
conjugate field is a vector potential\cite{Vignale1}.
 TDCDFT has since been applied to the calculation of the optical
spectra of solids \cite{deBoeij} and the polarizability of
long polymer chains \cite{Faassen} with considerable success.

In this Letter we show that the nonlocality problem occurs in
an {\it aggravated form} in the time-dependent spin density functional
theory or, more generally, in the time-dependent DFT of any multi-component
system.   The novel features of  the spin-dependent problem stem from the
fact that the xc kernel presents a divergence even
in the homogeneous electron liquid.  More precisely, it can be shown
that the Fourier transform of the spin-dependent exchange-correlation
kernel $f_{xc,\sigma \sigma'}(r-r',t-t') \equiv
\frac {\delta V_{xc,\sigma}(\vec r,t)}{\delta n_{\sigma'}(\vec r',t')}$ in
a homogeneous electron liquid has the long-wavelength expansion
\begin{equation} \label{fxcexpansion}
 f_{xc,\sigma \sigma'}(k,\omega) \stackrel{k \to 0} {\to}
\frac{A(\omega) }{ k^2} \frac{ \sigma \sigma'  n^2}{ 4n_\sigma n_{\sigma'}}
+B_{\sigma \sigma'}(\omega) + O(k^2)~,
\end{equation}
where $A(\omega)$ and $B_{\sigma \sigma'}(\omega)$ are
complex functions of frequency, $n_\sigma$ is the density
of $\sigma$-spin electrons ($\sigma = +1$ for $\uparrow$-spin  and
$\sigma = -1$ for $\downarrow$-spin),
and $n=n_\uparrow+n_\downarrow$ is the total density.
Since the xc potential  created by a small density variation
$\delta n_\sigma(\vec k, \omega)$ is given by the formula
\begin{equation} \label{Vxc}
V_{xc,\sigma}(\vec k, \omega) = \sum_{\sigma '}f_{xc,\sigma \sigma'}(k,\omega)
\delta n_{\sigma '}(\vec k, \omega)~,
\end{equation}
we see that Eq.~(\ref{fxcexpansion}) rules out the possibility
of a local connection between $V_{xc,\sigma}({\vec r}, t)$ and $\delta
n_{\sigma
'}({\vec r}', t')$.

The existence of the long-wavelength singularity in $f_{xc, \sigma \sigma'}
(k , \omega)$ has been known for some time.  It was first pointed out by
Goodman and Sj\"olander \cite{Goodman} that the third
moment sum rule for the spin-density response function implies such a
singularity.
Approximate formulae for $f_{xc,-} (k, \omega) = f_{xc, \uparrow \uparrow}
(k , \omega)
- f_{xc, \uparrow \downarrow} (k , \omega)$ exhibiting the singularity were
proposed in \cite{Liu} and  \cite{Richardson}.  More recently, D'Amico and
Vignale (\cite{DAmico}) have shown that, at low frequency and finite
temperature, the singularity is related to the friction that arises between up- and down-spin currents when they have different average velocities (the so-called spin-drag effect).

By contrast, the implications of Eq.~(\ref{fxcexpansion}) for spin density
functional theory have not been explored so far. This is understandable.
The singularity~(\ref{fxcexpansion}) arises only at finite frequency
($A(0)=0$) and therefore does not affect the {\it static} spin DFT.   Furthermore, the singularity does not show up as long as
one is interested only in the density response of spin-compensated systems,
since, in that case, the  relevant combination of xc kernels is $\sum_{\sigma
\sigma'}n_\sigma n_{\sigma '} f_{xc,\sigma \sigma'}$, which is non-singular.   It is only in the time-dependent  spin DFT \cite{Vosko} that the issue of the
singularity becomes really critical  not only to the calculation of the
spin response, but even to the calculation of just the density response\cite{footnote2}.

In this Letter we propose a resolution of the nonlocality problem
based on the use of the spin components
of the current density $\vec j_{\uparrow}(\vec r,\omega)$ and
$\vec j_{\downarrow}(\vec r,\omega)$ as basic variables.
We provide an explicit expression for the spin-dependent
exchange-correlation  field $\vec E_{xc,\sigma}(\vec r,\omega)$ as a local
linear functional of the currents $\vec j_\sigma$.

The general method for upgrading from the density to the current-density
formulation is described in detail in Ref.~\cite{ullrichlong}, so we
mention only the essential steps here.
We introduce a spin-dependent xc vector potential $\vec A_{xc,\sigma}(\vec
k, \omega)$
(whose time-derivative, $i \omega \vec A_{xc,\sigma}(\vec k, \omega)=\vec
E_{xc}(\vec k,\omega)$,
  is the xc electric field), and notice that this is linearly
related to the currents in the following manner
\begin{equation} \label{defaxc}
A_{xc,\sigma}^{\alpha}(\vec k, \omega)
~=~  \frac{k^2 }{ \omega^2} \sum_{\sigma '}f_{xc,\sigma \sigma '}^{\alpha}
(\vec k, \omega)j_{\sigma '}^{\alpha}(\vec k, \omega)~,
\end{equation}
where the superscript $\alpha$ denotes the longitudinal ($\alpha = L$) or
transverse ($\alpha = T$) component of a vector relative to the direction
of $\vec k$.
It is not difficult to see that the {\it longitudinal} xc kernel
defined in this manner coincides with the xc kernel introduced in
Eq.~(\ref{fxcexpansion}).
The extra factor $ \frac{k^2 }{ \omega^2}$ in Eq.~(\ref{defaxc})
exactly cancels the small-$k$ singularity of $f_{xc}$, and leads to a
theory that  admits a  local approximation.
The imaginary part of the current xc
kernel $f_{xc,\sigma \sigma '}^{\alpha}(k,\omega)$ is expressed in terms of a
causal response function as follows:
\begin{eqnarray} \label{Imfxc}
&& \Im m  f_{xc,\sigma \sigma '}^{\alpha}(k,\omega)
= \frac{1 }{ V   n_\sigma n_{\sigma '} k^2} \Im m \langle \langle  \hat
F_\sigma^\alpha(\vec k); \hat F_{\sigma'}^\alpha(-\vec k)
\rangle\rangle_\omega ~, \nonumber \\
\end{eqnarray}
where $\langle \langle \hat A;\hat B\rangle \rangle_\omega \equiv - \frac{
i}{ \hbar} \int_0^{\infty}\langle [\hat A(t),\hat B]  \rangle e^{i \omega
t}dt$ is the linear response function associated with the operators $\hat
A$ and $\hat B$;  $\hat F_\sigma^\alpha(\vec k) = -\frac{i m}{\hbar}[\hat H,
{\hat j}_{\sigma }^{\alpha} (\vec k)]$ is the time derivative of the
Fourier transform of the current-density operator  $\hat {\vec j}_\sigma
(\vec k)$, $\hat H$ is the Hamiltonian, and $V$ is the volume.

Once the imaginary part  of $f_{xc,\sigma \sigma
'}^{\alpha}(k,\omega)$ is known, its real part is determined by the Kramers-Kr\"onig dispersion
relation
\begin{eqnarray} \label{refxc}
\Re e  f_{xc,\sigma \sigma '}^{\alpha}(k,\omega)&=&
f_{xc,\sigma \sigma '}^{\alpha}(k,\infty)\nonumber  \\
&-& \frac{2 }{ \pi}{\cal P} \int_0^\infty d \omega '
\frac{\omega ' \Im m f_{xc,\sigma \sigma '}^{\alpha}(k,\omega ') }
{ \omega^2 - {\omega '}^2}~,\nonumber \\
\end{eqnarray}
where ${\cal P}$ denotes the principal part integral, and
the infinite frequency limit of  $f_{xc,\sigma \sigma '}$ is
determined by the {\it third moment sum rule}. In a three-dimensional
electron liquid, this sum rule gives
\begin{widetext}\begin{equation} \label{thirdmoment}
f_{xc,\sigma \sigma '}^{\alpha}(k,\infty) \stackrel{k \to 0}{\to}
-\frac{4 \pi e^2 }{ 3 k^2}\left [g_{\uparrow \downarrow}(0)-1 \right]\sigma
\sigma '  + \frac{1}{n_{\sigma}} \left [ a^\alpha    t_{c \sigma} \delta_{\sigma \sigma'} + b^\alpha
 \epsilon_{pot, \sigma \sigma'} \right ]
\end{equation} \end{widetext}
where $a^L=2$, $a^T=2/3$,  $b^L=4/15$, and $b^T=-2/15$.
Here $g_{\uparrow \downarrow}(0)$ is the pair correlation function for
antiparallel spin electrons at zero separation,  $t_{c \sigma}$ is the
average correlation kinetic energy of the $\sigma$-spin component, and $
\epsilon_{pot, \sigma \sigma'}  \equiv \frac{n_\sigma}{2} \int d \vec r \frac {e^2}{r} [g_{\sigma \sigma'}(r)-1]$ is the potential energy 
associated with the interaction between $\sigma$- and $\sigma'$-spin
electrons. Note that the result for the longitudinal case was first
obtained in Ref.\cite{Goodman}.

It is evident from the above equations that both  the longitudinal and
the transverse kernels exhibit  $\frac{1 }{ k^2}$ singularities,
which are ``cured" by the $\frac{k^2 }{ \omega^2}$ factor of
Eq.~(\ref{defaxc}).
In particular, substituting  the  small-$k$ expansion $\hat
F_{\sigma}^\alpha (\vec k) = \hat F_\sigma^\alpha(0)+ O(\vec k)$ in
Eq.~(\ref{Imfxc}),  where $\hat F_\sigma^\alpha(0)$ is the operator of the
total force acting on $\sigma$-spin electrons, and noting that terms of
first order in $\vec k$ vanish by inversion symmetry, we see that the
xc-kernels have the small-$k$ expansion
\begin{equation} \label{fxccurrentexpansion}
 f_{xc,\sigma \sigma'}^{\alpha}(k,\omega) \stackrel{k \to 0}{\to}
\frac {A(\omega) }{ k^2}\frac{ \sigma \sigma'  n^2}{ 4n_\sigma n_{\sigma'}}
+B_{\sigma \sigma'}^\alpha(\omega) + O(k^2)~,
\end{equation}
where
\begin{equation}\label{imaxc}
\Im m A(\omega) = - \frac{4}{V n^2} \Im m \langle \langle \hat
F^\alpha_\uparrow;\hat F^\alpha_\downarrow\rangle \rangle_\omega~,
\end{equation}
and
\begin{eqnarray} \label{reaxc}
\Re e  A(\omega) &=&
-\frac{16 \pi e^2 }{ 3}\left [g_{\uparrow \downarrow}(0)-1 \right]
\nonumber \\
&-& \frac {2 }{ \pi}{\cal P} \int_0^\infty d \omega '
\frac {\omega ' \Im m A(\omega ') }{ \omega^2 - {\omega '}^2}~.
\end{eqnarray}
The factor $\sigma \sigma'$ in Eq.~(\ref{fxccurrentexpansion}) arises from the
fact that the total force $\hat F_\uparrow + \hat F_\downarrow$ vanishes by
translational
invariance, so that $\langle \langle \hat F_\sigma;\hat F_{\sigma'}\rangle
\rangle_\omega
= - \sigma \sigma ' \langle \langle \hat F_\uparrow;
\hat F_{\downarrow}\rangle \rangle_\omega$.
Notice also that $A(\omega)$ is independent of the direction $\alpha$ -
longitudinal or transverse.
The microscopic expression for $B^\alpha_{\sigma \sigma'}$ is
more complicated:  a simple approximation for this quantity will be presented below.

Substituting the expansion~(\ref{fxccurrentexpansion}) in
Eq.~(\ref{defaxc}), calculations similar to those described in
~\cite{ullrichlong} lead us to the following {\it local}
approximation for the xc field in terms of the spin currents
\begin{eqnarray} \label{localexc}
-e \vec E_{xc,\sigma}(\omega) &=& -\vec \nabla V_{xc,\sigma}^{LDA}
+\frac{1}{n_{\sigma} }\vec \nabla \cdot {\bf \stackrel{\leftrightarrow}{\sigma}}_{xc,\sigma}(\omega)
\nonumber \\
&+& \frac{i n^2 A (\omega)}{4 \omega}
\sum_{\sigma'}
\frac{\sigma \sigma'}{ n_\sigma n_{\sigma'}} \vec j_{\sigma'}~.
\end{eqnarray}
Here the $\vec r$ dependence has been left implicit,
and the xc stress tensor ${\bf \stackrel{\leftrightarrow}{\sigma}}_{xc}(\omega)$, as well as $A(\omega)$, is a function of the local spin densities, as discussed below.

Eq.~(\ref{localexc}) is the central result of this paper.  The first two terms
on the right are well known: they are, respectively,  the adiabatic LDA
contribution
and the visco-elastic force term, where the  stress tensor
$\sigma_{xc,\sigma}(\omega)$
is related to $B_{xc,\sigma \sigma'}$ by obvious extensions of the formulae
reported in ~\cite{ullrichlong}. The expression for the xc stress tensor is
\begin{eqnarray}
\sigma_{xc, \sigma, ij} &=& \sum_{\sigma'}
\biggl [ \eta_{xc, \sigma \sigma'} \biggl (
\frac{\partial u_{\sigma', i}}{\partial r_j}
+ \frac{\partial u_{\sigma', j}}{\partial r_i}
- \frac{2}{3} \vec \nabla \cdot {\vec u}_{\sigma'} \delta_{ij} \biggr )
\nonumber \\
&+& \zeta_{xc, \sigma \sigma'} \vec \nabla \cdot {\vec u}_{\sigma '}
\delta_{ij} \biggr ]
\end{eqnarray}
where $\vec u_{\sigma} = \vec j_{\sigma} / n_\sigma $, and
\begin{eqnarray}
\eta_{xc, \sigma \sigma '} = - \frac{n_\sigma n_{\sigma '}}{i \omega}
B_{\sigma \sigma '}^T (\omega )~,
\end{eqnarray}
\begin{eqnarray}
\zeta_{xc, \sigma \sigma '} = - \frac{ n_\sigma n_{\sigma '}} {i \omega }
[ B_{\sigma \sigma '}^{L} (\omega) - \frac{4}{3} B_{ \sigma \sigma '}^{T}
(\omega)
- \epsilon_{xc, \sigma \sigma '}'' ]~,
\end{eqnarray}
where $\epsilon_{xc, \sigma \sigma '}'' = \frac{\partial^2
\epsilon_{xc}}{\partial n_\sigma \partial n_{\sigma'}}$.
The last term in Eq.~(\ref{localexc}) is new, and comes
directly from the $\frac{1}{k^2}$ singularity of
Eq.~(\ref{fxccurrentexpansion}).
%In fact, one can easily show that $\gamma (\omega)$ is related to
%$A(\omega)$ by
%\begin{equation} \label{defgamma}
%\gamma(\omega) = - \frac {i n^3 A(\omega)}{4 \omega m n_\uparrow
%n_\downarrow}~.
%\end{equation}
The essential feature of the new term is that it produces
damping of the spin current proportional to the relative velocity between
up- and down-spin electrons.  This makes it readily distinguishable from
the usual viscous friction contained in the second term,  which is
proportional
to the {\it derivatives} of the velocity field.
\begin{figure} \label{fig1}
\includegraphics[width=7cm]{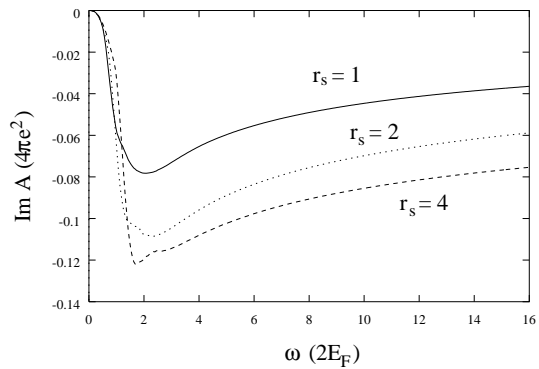}
\caption {Imaginary part of $A(\omega)$ evaluated from Eq.~(\ref{ImA}) with
the correction factor given in Eq.~(\ref{factor}).  The values of $a(r_s,
0)$ are 1.92, 3.36, and 7.49 at $r_s=1,2$, and $4$ respectively.}
\vspace{10pt}
\end{figure}
\begin{figure} \label{fig2}
\includegraphics[width=7cm]{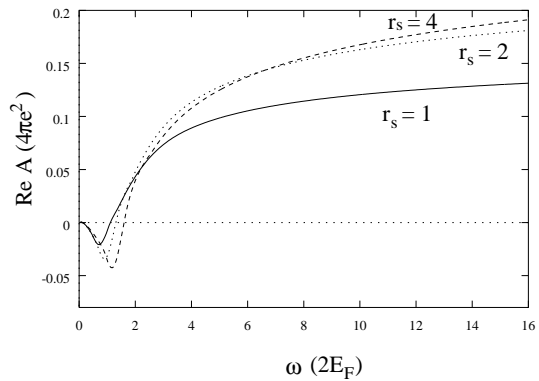}
\caption {Real part of $A(\omega)$ calculated from Eq.~(\ref{reaxc}).}
\vspace{0pt}
\end{figure}
The physical reason for the difference is that, whenever up and down spin
currents travel with different average velocities, they exert {\it
friction} on each other:  the ``spin drag coefficient" is $\gamma (\omega
) = \frac{i n^3 A(\omega )}{4 \omega m n_\uparrow n_\downarrow}$.
Of course, like all the quantities considered here, $\gamma(\omega)$ is complex
and frequency-dependent, and, in the limit of zero frequency, its real part
can be shown to be related to the spin diffusion constant $D_s$ by the
Einstein relation $D_s = \frac {n}{m \chi_s \gamma(0)}$, where $\chi_s$
is the static, macroscopic spin susceptibility.

Unfortunately, an exact calculation of $A(\omega)$ from the microscopic
expressions~(\ref{imaxc})
and (\ref{reaxc}) is beyond the reach of present-day many-body techniques.
However, we can obtain a rather good approximation with the help of the
following
exact results:  (i) For $\omega \to 0$, $\Im m A(\omega) \propto \omega^3$
and $\Re e A(\omega) \propto \omega^2$;
(ii) For large $\omega$, $\Im m A(\omega) \to -\frac{16 \pi e^2}{3}
\frac {n_\uparrow n_\downarrow}{n^2} \frac{\alpha r_s}{\sqrt{\bar \omega}}
\frac{1}{(1+\zeta)^{1/3}}$ and $\Re e A(\omega) \to -\frac{16 \pi e^2}{3}
\frac {n_\uparrow n_\downarrow}{n^2} [g_{\uparrow \downarrow}(0)-1]$. Here $\bar \omega = \frac{\omega}{2 E_{F \uparrow}}$, where $ E_{F
\uparrow}$ is
the Fermi energy for majority spin electrons and
$\zeta = \frac{n_{\uparrow}-n_{\downarrow}}{n}$ measures the degree of spin polarization, and $\alpha=(4/9 \pi)^{1/3}$ \cite{footnote3}.
Note that $g_{\uparrow \downarrow}(0)$ is accurately known from
the work of Gori-Giorgi and Perdew \cite{Perdew}.
The high and low frequency limits  of $\Re e A(\omega)$ are both obtained
from the third moment sum rule.  In particular, the vanishing of $\Re e A(0)$
follows from the fact that $\frac{2}{\pi}\int_0^\infty \frac {\Im m
A(\omega')}{\omega'}$
is equal to (minus) the first moment of the current-current response function,
which, by gauge invariance and the continuity equation,
coincides with the third moment of the density-density response function,
i.e., $-A(\infty)$.

The $\omega^3$ behavior of $\Im m A(\omega)$ at low frequency is easily obtained from the approximate zero-temperature formula \cite{DAmico}
\begin{widetext}
\begin{eqnarray} \label{ImA}
\Im mA(\omega)  \simeq  - \frac {4}{3 n^2 V}\sum_{\vec q}v_{\vec q}^2 q^2
\int_0^\omega \frac{d \omega'}{\pi} \left [\Im m \chi_{\uparrow \uparrow}
(q,\omega - \omega')\Im m \chi_{\downarrow \downarrow}(q,\omega')
- \Im m \chi_{\uparrow \downarrow}(q,\omega - \omega')
\Im m \chi_{\downarrow \uparrow}(q,\omega') \right ]~,
\end{eqnarray}
\end{widetext}
which is exact in the limits of
 high density  and high frequency. Here  $v_{\vec q}=\frac{4 \pi e^2}{q^2}$, and $\chi_{\sigma \sigma'}(q,\omega)$ are the spin density
response functions of the homogeneous liquid.

We have evaluated $\chi_{\sigma \sigma'}$in the generalized random phase approximation
\begin{equation}
\chi^{-1}_{\sigma \sigma'}(q,\omega)=\chi^{-1}_{0\sigma}(q,\omega) \delta_{\sigma \sigma'}-v_{\vec q}[1-G_{\sigma \sigma'}(q)]~,
\end{equation}
where $\chi_{0\sigma}(q,\omega)$ is the Lindhard function and $G_{\sigma \sigma'}(q)$ are local field corrections \cite{Iwamoto}.
At  typical metallic densities we multiply $ \Im m A(\omega)$   by an empirical factor
\begin{equation} \label{factor}
g(\omega)=\frac{1 + \sqrt{\bar \omega}}{a(r_s, \zeta) +
\sqrt{\bar \omega}}~,
\end{equation}
designed to satisfy the sum rule $\Re e A(0) = 0$
without altering the high-frequency behavior. Notice that $a(r_s, \zeta)
\to 1 $ for $r_s \to 0 $.  The results evaluated with this procedure are shown in Figs. 1 and 2.
 %An alternative approach to the calculation of $A(\omega)$ is to assume,
%in the spirit of the Gross-Kohn approximation (\cite{GK}), the approximate
%form
%\begin{equation}
%\frac{\Im m A(\omega)}{4 \pi e^2} = \frac{a \bar \omega^3}
%{\left (1 + b \bar \omega^2 \right)^{7/4}}~,
%\end{equation}
%which satisfies all the limiting forms provided that $a = ....$ and $b =
%A comparison between this simple interpolation and the mode coupling
%calculation is presented in Figures 1 and 2.

Finally, we briefly remark on the calculation of the regular
part $B^\alpha_{\sigma \sigma'}(\omega)$
of $f^{\alpha}_{xc,\sigma \sigma'}$.
A simple approximation strategy is as follows.
We rewrite the $2 \times 2$ matrix $f^{\alpha}_{xc,\sigma \sigma'}$
in the basis of the vectors $|r\rangle \propto  (n_\uparrow, n_\downarrow)$
and $|s \rangle \propto  (-n_\downarrow, n_\uparrow)$.  It is immediately
seen that only the matrix element $f^{\alpha}_{xc,rr}$ is
finite in the limit $k \to 0$, while all the others are singular.  This
suggests
that we approximate $f^{\alpha}_{xc,rr} \sim B^\alpha_{rr}$   and completely
ignore the contribution from $B^\alpha_{\sigma \sigma'}(\omega)$ in all other matrix elements, which are
dominated by the singular contribution of $A(\omega)$.  A little thought shows
that this approximation is equivalent to setting
\begin{equation}
 B^\alpha_{\sigma \sigma'}(\omega) \simeq  \frac{n_\sigma n_{\sigma'}}{n^2}
B^\alpha (\omega)~,
\end{equation}
where the scalar function $B^\alpha (\omega)$   can be extracted from
a calculation of  the xc kernels in the density channel.  Such a calculation
has been carried out in \cite{QV} (for the paramagnetic state) and
the connection between $B^\alpha (\omega)$ and the $f_{xc}^\alpha (\omega)$
of that paper
is $B^\alpha(\omega) = \frac{f_{xc}^\alpha (\omega)}
{\left ( 1 - 2 \frac{n_\uparrow n_\downarrow}{n^2} \right )^2}$.

This completes the construction of the input  necessary to the evaluation of Eq.~(\ref{localexc}).  It is hoped that our expression for
the spin-current dependent xc field ~(\ref{localexc}) will open the way to novel applications of CDFT to the calculation of spin excitations in spin-polarized systems.  

This work was supported by NSF grant No. DMR-0074959.  We acknowledge useful discussions with Carsten Ullrich.


\begin{thebibliography}{100}
\bibitem{footnote1} More precisely, the Fourier transform of the xc kernel $f_{xc}(\vec k, \vec k')$ must have a finite limit for $\vec k$ and/or  $\vec k'$ tending to zero.
\bibitem{footnote3} The large $\omega$ limit of $\Im m A (\omega)$
at $\zeta =0$ was also obtained by Liu \cite{Liu}. However, his
assumption that $\Im m A (\omega) \propto \omega$ for $\omega \to 0$ is incorrect.
\end{thebibliography}
\end{document}